\documentstyle{mn}
\input epsf
\newcommand{\reff}{\mbox{$R_{\rm eff}$}}

\newcommand{\mto}{\mbox{${\rm M}^{\rm TO}_V$}}
\newcommand{\vto}{\mbox{${\rm m}^{\rm TO}_V$}}
\newcommand{\ishape}{{\sc ishape}}

\newcommand{\tinytim}{{\sc TinyTim}}

\newcommand{\bvz}{\mbox{$(\bv)_0$}}
\newcommand{\bv}{\mbox{$B\!-\!V$}}

\newcommand{\vi}{\mbox{$V\!-\!I$}}

\newcommand{\viz}{\mbox{$(V\!-\!I)_0$}}

\newcommand{\apj}{ApJ}

\newcommand{\aap}{A\&A}
\newcommand{\aaps}{A\&AS}
\newcommand{\pasp}{PASP}
\newcommand{\aj}{AJ}
\newcommand{\araa}{ARA\&A}

\newcommand{\mnras}{MNRAS}

\begin{document}
\title[Globular clusters in the Sombrero]{HST photometry of globular 
clusters in the Sombrero galaxy
\thanks{Based on observations with the NASA/ESA Hubble Space
 Telescope, obtained at the Space Telescope Science Institute, which is
 operated by the Association of Universities for Research in Astronomy,
 Inc. under NASA contract No.  NAS5-26555.}
}
\author[S. S. Larsen et al.]
  {S{\o}ren S. Larsen,$^1$ Duncan A. Forbes$^2$ and Jean P. Brodie$^1$ \\
  $^1$UC Observatories / Lick Observatory, University of California 
       Santa Cruz, CA 95064, USA 
       email: soeren@ucolick.org \\
  $^2$Astrophysics \& Supercomputing, Swinburne University, Hawthorn VIC 
3122, Australia}

\maketitle
\begin{abstract}
We explore the rich globular cluster (GC) system of the nearby Sa galaxy 
M104, the ``Sombrero'' (NGC~4594), using archive WFPC2 data.  The GC colour 
distribution is found to be bimodal at the $>99\%$ confidence level, with 
peaks at $(V-I)_0 = 0.96\pm0.03$ and $1.21\pm0.03$. The inferred metallicities 
are very similar to those of globular clusters in our Galaxy and M31. 
However the Sombrero reveals a much enhanced number of red (metal-rich) 
GCs compared to other well-studied spirals. Because the Sombrero is dominated
by a huge bulge and only has a modest disk, we associate the two 
sub-populations with the halo and bulge components respectively.  Thus our 
analysis supports the view that the metal-rich GCs in spirals are associated 
with the bulge rather than with the disk.  The Sombrero GCs have typical 
effective (half-light) radii of $\sim$2 pc with the red ones being $\sim$30\% 
smaller than blue ones.  We identify many similarites between the Sombrero's GC system 
and those of both late type spirals and
early-type galaxies. Thus both the GC system and the Hubble type of the 
Sombrero galaxy appear to be intermediate in their nature.
\end{abstract} 


\section{Introduction}

  Observations with the {\it Hubble Space Telescope} have, in many ways,
revolutionized our understanding of globular cluster systems (GCSs) in external
galaxies. However, ellipticals and, to some extent, S0 galaxies have received 
by far the largest amount of attention, primarily because early-type galaxies
tend to have much richer GCSs and suffer less from internal extinction 
problems than spirals. As a result, our knowledge of the GCSs in 
{\it spiral} galaxies is still limited to a handful of galaxies, 
including the Milky Way and M31.

  Perhaps the most conspicuous difference between the GCSs of ellipticals
and spirals is the much higher specific frequencies ($S_N = $ number of 
globular clusters per unit galaxy luminosity, Harris \& van den Bergh 1981) 
of ellipticals.  Spiral galaxies typically have $S_N \la 1$, while normal 
ellipticals have $S_N \sim 2 - 5$ and cDs often reach $S_N\sim 15$ 
\cite{har91}. Part of this discrepancy may be removed if the $S_N$ of 
spirals is normalized only to the spheroidal component rather than
to the total galaxy luminosity \cite{cot00}, although this approach becomes 
problematic for galaxies like the LMC and M33 which lack an obvious 
spheroidal component but do possess significant populations of globular 
clusters (e.g.\ Forbes et al.\ 2000). McLaughlin \shortcite{mac99} has
argued that the high specific frequencies of giant ellipticals may be
explained if the number of globular clusters is instead normalised to the
total galaxy mass, including hot X-ray emitting gas. In this way he finds
a constant globular cluster formation efficiency with respect to the
total initial gas mass available, but the remaining problem is then to
explain why the \emph{field star} formation efficency in the giant ellipticals
is lower than that of GCs.  

  In addition to the Local Group galaxies, some of 
the best studied spirals are M81 and the two edge-on spirals NGC~4565 and 
NGC~5907. Although M81 is nearby, studies of its GCs are complicated by 
internal extinction in M81 and by its large angular size which causes
confusion problems with foreground stars.  However, available data indicate 
that the GCS of M81 is very similar to that of M31 and the Milky Way 
\cite{per95,sch01}. In NGC~4565 and NGC~5907, 
internal extinction problems are minimized because of the edge-on orientation 
of these two galaxies. Again, the GCSs of these two galaxies (type Sb and 
Sc, respectively) appear to be quite similar to those of the Milky Way and 
M31 in terms of richness and spatial structure, with specific frequencies 
of $S_N \sim 0.5$ \cite{fle95,kis99}. 

  It has been known for many years that the Milky Way globulars can 
be divided into (at least) two populations, based on their metallicities 
and spatial distributions \cite{kin59,zin85,min95}, where the {\it metal-rich} 
population appears to be associated with either the thick disk or the 
bulge component. Similar conclusions have been reached for the GCS
of the Andromeda galaxy \cite{sei96,jab98,barm00}. In contrast, Kissler-Patig 
et al. \shortcite{kis99} suggested that, since NGC~4565 and NGC~5907 
contain roughly equal numbers of globular clusters but have very different 
bulge to thin-disk and thick-disk to thin-disk ratios, these components of 
spiral galaxies may {\it not} have a significant influence on the building 
of globular cluster systems around such galaxies.  However, metal-poor and 
metal-rich clusters were not addressed separately in the study of NGC~4565 
and NGC~5907 and indeed the total numbers of GCs detected in these galaxies 
(40 and 25 respectively) were too small to permit such an analysis.

  Within the last few years it has been firmly established that the GCSs
of many, if not most, early-type galaxies also exhibit bimodal metallicity 
distributions \cite{geb99,kundu01,lar01}. The spatial distribution of red 
(metal-rich) clusters tends to be more centrally concentrated than that of
the blue (metal-poor) ones. 
One of the more surprising results of
recent HST imaging has been the discovery that the average {\it sizes} of 
metal-poor and metal-rich GCs in early-type galaxies differ by 
$\sim 30$\%, with the metal-poor clusters being systematically larger
\cite{kundu98,kundu99,puz99,lar00}.  The sizes of globular clusters in 
the Milky Way also show a correlation with galactocentric distance, in 
the sense that clusters far from the centre are systematically larger 
\cite{van94}. This appears to be in contrast to the situation in elliptical 
galaxies, at least in  NGC~4472 and M87 where no such trend is seen 
\cite{kundu99,puz99}. Furthermore, the size difference is apparently 
present over a large range of galactocentric distances \cite{lar01}, so 
at least in some ellipticals it seems difficult to explain it as simply due 
to differences in the radial distributions of metal-rich and metal-poor 
globular clusters.

  The ``Sombrero'' galaxy (M104 = NGC~4594) represents an important 
intermediate case between early-type galaxies and the Sb/Sc type spirals 
mentioned above. Although it clearly contains a star forming disk and a 
conspicuous dust lane, its integrated luminosity is dominated by 
a huge bulge/halo component. It is one of the nearest Sa--type galaxies, 
located at a distance of only 8.7 Mpc \cite{for96}.  The first study of 
globular clusters in the Sombrero was carried out by Wakamatsu 
\shortcite{wak77} who identified a very rich GCS with an estimated total of 
about 2000 GCs.  Harris et al. \shortcite{har84} identified $1200\pm100$ 
globular clusters in the Sombrero, corresponding to a specific frequency of 
$3\pm1$ with respect to the bulge/halo luminosity alone. A similar number of 
GCs was found by Bridges \& Hanes \shortcite{bri92} who also estimated a 
mean metallicity
of [Fe/H] = $-0.81\pm0.05$ based on \bv\ colours. This mean metallicity was
confirmed spectroscopically by Bridges et al. \shortcite{bri97} who obtained
a mean [Fe/H] = $-0.70\pm0.3$ for a sample of 34 GCs.  Ground-based CCD 
imaging of the GCS by Forbes et al. \shortcite{for97a} found some 
tentative evidence for two populations of GCs.  However these results were
uncertain due to contamination from foreground stars and background galaxies 
in their ground-based images. Two populations were confirmed by Gebhardt \& 
Kissler-Patig \shortcite{geb99} in their study of 50 galaxies using 
archive WFPC2 data.  However they did not comment on the Sombrero 
galaxy explicitly. 

  Here we present a new investigation of the globular cluster system of
the Sombrero galaxy, using HST archive images of three fields.  

\begin{table*}
\caption{\label{tab:hstpt}
  WFPC2 archive images of the Sombrero galaxy used in this paper.
}
\begin{tabular}{lclcc} \hline
Field & PID & PI & Filters & Exposures \\
Centre     & 5512 & Faber & F547M $+$ F814W & $3\times400 + 3\times350$ s \\
Inner Halo & 5091 & Groth & F606W $+$ F814W & $800 + 800$ s \\
Outer Halo & 5369 & Griffiths & F606W $+$ F814W & $2\times 2100 + 2\times 2100$ s \\ 
Reference  & 7909 & Casertano & F606W $+$ F814W & $500 + 400,500$ s \\ \hline
\end{tabular}
\end{table*}


\section{Data reduction}

\begin{figure}
\epsfxsize=85mm
\epsfbox{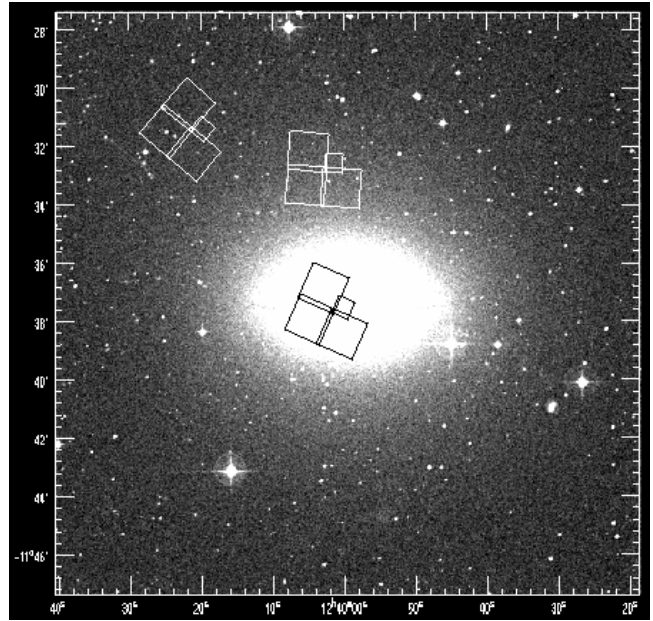}
\caption{\label{fig:hstpt} The three WFPC2 pointings on the Sombrero
galaxy used for this study, superposed on an image from the
Digitized Sky Survey.
}
\end{figure}

  Basic information about the datasets is listed in Table~\ref{tab:hstpt} 
and the pointings are illustrated in Fig.~\ref{fig:hstpt}.  The central
pointing has the PC chip centred on the nucleus of the Sombrero galaxy and
consists of $3\times400$ s exposures in the F547M band and 
$3\times350$ s in F814W. The WFPC2 field reaches out to a distance of
about 4 kpc from the nucleus.  The two halo pointings were both observed
with the F606W and F814W filters and are located at $4\farcm6$
(11 kpc projected) and $8\farcm3$ (21 kpc projected) from the centre of 
the Sombrero.  Unfortunately, only one 
exposure of 800 s in each filter was available for the inner halo pointing, 
which makes cosmic ray removal a more difficult task.  For the outer 
pointing we use 2 exposures of 2100 s each in each filter, far deeper than 
the two other pointings. In addition, we use F606W and F814W data for a 
reference field at $227\arcmin$ from the Sombrero to estimate the number
of foreground/background sources.

  For the pointings where more than one exposure was available in each
filter, the individual exposures in each band were combined into a single 
image using the 
IRAF\footnote{IRAF is distributed by the National Optical Astronomical 
Observatories, which are operated by the Association of Universities 
for Research in Astronomy, Inc.~under contract with the National 
Science Foundation}
task {\sc imcombine}. For the central pointing, 
background subtraction was then performed by first removing point sources 
and then median filtering the resulting images as described in 
Larsen \& Brodie \shortcite{lar00}. The median-filtered images were
then subtracted from the original images. 

  The region around the dust lane and the high surface brightness areas
near the nucleus were masked out before further analysis. Input object lists 
for photometry were then produced by running the {\sc daofind} task in 
the {\sc daophot} package within IRAF on both the $V$ and $I$ band images and 
matching the two resulting coordinate lists. As an additional selection 
criterion, the 
background noise was measured directly on the images in an annulus around 
each object and only objects with a S/N $>3$ in both $V$ and $I$ within an 
aperture radius of 2 pixels were accepted. In this way we accounted for 
the varying background noise, especially within the central pointing. 
For the inner halo pointing where the individual exposures suffered from
quite substantial numbers of cosmic ray (CR) events, the object lists were
inspected manually to make sure that all sources for which photometry
was subsequently obtained would be real objects. The requirement that 
objects were detected both in $V$ and $I$ turned out to be quite efficient
in eliminating CR hits from the object lists. However, a few GC candidates
were manually removed from the object lists because of CR hits that would
have affected the photometry.

  The \vi\ colours were measured through an $r=2$ pixels aperture using 
the {\sc phot} task in {\sc daophot}, while a larger $r=3$ pixels
aperture was used for $V$ magnitudes to reduce systematic errors
on the aperture corrections resulting from finite object sizes. The use
of a smaller aperture for colours is justified because the $V$ and $I$
aperture corrections nearly cancel out when colour indices are
formed \cite{hol96,lar00}.  The photometry was
transformed to standard $V, I$ magnitudes using the procedure in
Holtzman et al. \shortcite{hol95}.  Globular clusters are expected 
to be resolved on HST images at the distance of M104 so we determined
aperture corrections to the Holtzman et al. $r=0\farcs5$ reference aperture 
by convolving King profiles with the \tinytim\ PSF \cite{kri97} and 
carrying out 
aperture photometry on the resulting images. Because of the different
image scales on the WF and PC chips, the aperture corrections will also be
different. We found aperture corrections in $V$ of $-0.15$ mag and $-0.55$ 
mag for the WF and PC, respectively. These corrections assume clusters with 
effective (half-light) radii $\reff \sim 3$ pc. For larger clusters
our magnitudes will be systematically too faint, by $\sim 0.2$ mag on
the WF and $\sim 0.3$ mag on the PC chip if the cluster sizes are twice 
as large. For \vi\ the aperture corrections are $0.030$ and 
$0.120$ mag for the WF and PC, changing by no more than $\sim0.01$ mag 
for any reasonable cluster sizes. Finally, a correction for Galactic
foreground extinction of $A_V = 0.17$ mag \cite{sch98} was applied to
the photometry, corresponding to $E(V-I) = 0.086$ \cite{car89}.

  In addition to the photometry, we also measured sizes for individual
objects in the HST images using the \ishape\ algorithm \cite{lar99a}.
\ishape\ convolves the PSF (in this case generated by \tinytim ) with
a series of different model profiles, adjusting the size (FWHM) of the 
model until the best possible match to the observed profile is obtained.
In the case of HST images, \ishape\ also convolves with the WFPC2
``diffusion kernel'' \cite{kri97}. For the model profiles we adopted
King \shortcite{king62} profiles with a concentration parameter 
(ratio of tidal vs.\ core radius) of 30.  Fitting the concentration parameter 
and linear size simultaneously would require better signal-to-noise and/or 
angular resolution than what is currently available for the Sombrero.
However, as discussed in Larsen \shortcite{lar99a}, the effective radii 
derived from the \ishape\ fits are not very sensitive to a particular choice 
of model profile as long as the sources have intrinsic sizes smaller than 
or comparable to that of the PSF.

\section{Results}

\subsection{The central pointing}

  We first consider the data for the central pointing, as this is
where the largest number of globular clusters was found.

\begin{figure}
\epsfxsize=85mm
\epsfbox{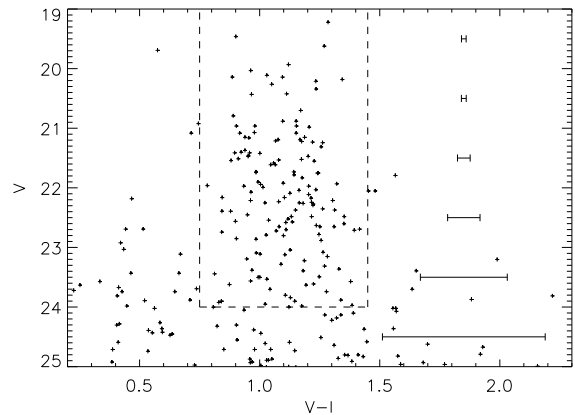}
\caption{\label{fig:cmd} Colour-magnitude diagram for objects
in the WFPC2 chips on the central pointing. The dashed lines mark the 
boundary of the region within which globular cluster candidates were 
selected. Typical \vi\ errors are indicated by the error bars.
}
\end{figure}

\begin{figure}
\epsfxsize=85mm
\epsfbox{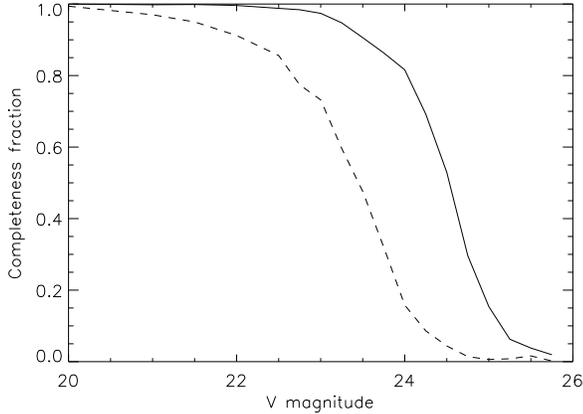}
\caption{\label{fig:cmpl} Completeness functions for the WF frames (solid
line) and PC frame (dashed line), determined from artificial object
experiments.
}
\end{figure}

  Fig.~\ref{fig:cmd} shows the reddening corrected colour-magnitude 
diagram for all objects detected in the central pointing down to $V=25$. 
The globular cluster sequence can be clearly seen, extending 
from $V\sim 20$ down to $V\sim24$ and with \vi\ colours between 0.8 and 1.4.  
Objects outside this colour range are expected to be mainly background 
sources. We selected globular cluster (GC) candidates as objects within 
the region marked by the dashed lines in Fig.~\ref{fig:cmd}, i.e.  
$0.75<\viz < 1.45$ and $V<24$.  This results in the identification of 
151 GC candidates, listed in Table~\ref{tab:clusters}. If we convert the
colour interval to a metallicity range using the calibration of Kissler-Patig 
et al. \shortcite{kis98}, then it corresponds to $-2.05 < \mbox{[Fe/H]} 
< +0.24$, but we note that these limits should only be taken as approximate 
since they are beyond the actual calibrated range.

  Completeness tests for the photometry were carried out by adding artificial
objects to the images and redoing the photometry to see how many of the 
artificial objects were recovered. Because the completeness functions will 
depend on object size, we generated the artificial objects by convolving the 
\tinytim\ PSF with King profiles with effective (half-light) radii of 3 pc.  
Since the GC candidates do not show a strong concentration towards the centre
of the galaxy and we are masking out the regions near the dustlane, the 
artificial objects were simply distributed at random 
within each frame.  The recovery fractions are shown as a function of 
magnitude in Fig.~\ref{fig:cmpl} for the WF (solid line) and PC frames (dashed 
line).  From these tests we estimate the average 50\% completeness level to 
be at $V\sim 24.5$ in the WF frames and at $V\sim 23.5$ in the PC frame.  The 
brighter 50\% limit in the PC frame is mainly due to the higher background 
level there. We thus reach well below the expected turn-over of the globular 
cluster luminosity function (GCLF) at $M_V \sim -7.5$ \cite{har91}, 
corresponding to $V\sim22.4$.  

  The number of contaminating objects can be estimated from the 
comparison field. This field is located so far from the Sombrero galaxy
($227\arcmin \sim 575$ kpc) that no globular clusters are expected here. 
We find about 8 objects in this field resembling globular clusters by
their colours and morphological appearance.
Some of the objects in 
the comparison field are unresolved by \ishape\ (presumably foreground stars)
while others have apparent sizes comparable to globular clusters. The latter 
are most likely background galaxies.

\subsubsection{Colour distribution}

  A glance at Fig.~\ref{fig:cmd} suggests two 
peaks in the \viz\ colour distribution. Applying a KMM test \cite{ash94} to 
the sample of GC candidates, we find that the colour distribution is 
bimodal at the 99.9\% confidence level with peaks at \viz\ = 0.96 and 
1.21.  These \vi\ colours correspond to metallicities of [Fe/H] = $-1.4$ 
and [Fe/H] = $-0.54$ when using the relation of Kissler-Patig et al.  
\shortcite{kis98}.  The KMM test itself does not provide error estimates 
for the peak colours but from experiments with different selection criteria 
for GC candidates we find that the peak colours are probably accurate to 
about $\pm 0.03$ mag, corresponding to $\pm 0.1$ in [Fe/H].
The KMM test assigns 45\% of the GC candidates in the central pointing to 
the blue (metal-poor) peak and 55\% to the red one.

\begin{figure}
\epsfxsize=85mm
\epsfbox{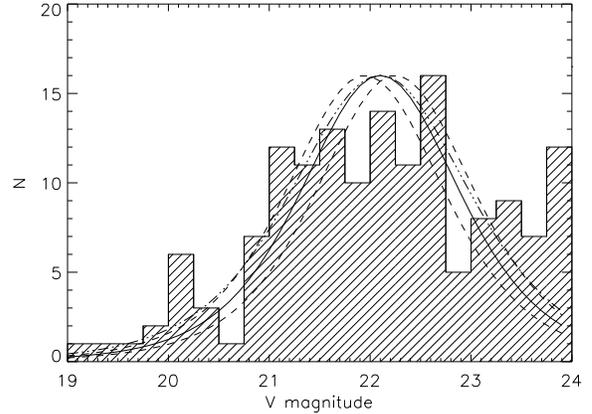}
\caption{\label{fig:gclf} Luminosity function for globular cluster
candidates in the NGC~4594 central pointing. The solid line shows
the best fitting $t_5$ function, $\mto = 22.10$ and $\sigma_t=0.81$. 
Dashed lines: $t_5$ functions with the turn-over shifted by 
$\pm0.15$ mag (the estimated $1\sigma$ error).  Dotted-dashed line: 
$\sigma_t = 0.91$.
}
\end{figure}

\subsubsection{Luminosity function}

  The luminosity function (LF) for GC candidates in the central pointing
(corrected for Galactic foreground extinction) is shown in Fig.~\ref{fig:gclf}. 
The LF clearly has a peak at $V\sim22$, although the histogram in 
Fig.~\ref{fig:gclf} has more faint objects than would be expected for a 
``standard'' Gaussian LF.  The field around the Sombrero galaxy has many
background galaxies and some of the faintest objects may be contaminants, 
but because the halo fields (see below) and the comparison field do not
contain the same large numbers of faint objects within the GC colour range,
part of the departure from a Gaussian LF may well be real. As noted 
e.g.\ by Secker \shortcite{sec92}, there are other analytical functions 
which actually match the faint end of the GCLF in the Milky Way and M31 
galaxies significantly better than a Gaussian, notably the `$t_5$' function. 
The $t_5$ and Gaussian functions are hardly distinguishable above the 
turn-over, but for magnitudes fainter than the turnover a $t_5$ function has 
larger numbers of objects than a Gaussian.

  We carried out $t_5$ function fits to the data in Fig.~\ref{fig:gclf} using 
a maximum-likelihood algorithm. A correction for completeness was applied, 
using the completeness function derived from the artificial object tests,
and contaminating objects were statistically subtracted from the GC sample 
using the comparison field. Objects in the PC frame were excluded from
the fits because of the brighter completeness limit there.  Fitting a $t_5$ 
function to all data in the WF chips down to $V=24$, we found a  peak at 
$\vto=22.28\pm0.12$ and a dispersion of $\sigma_{t} = 1.02\pm0.12$.
Note that the dispersions of $t_5$ ($\sigma_t$) and Gaussian ($\sigma_g$) 
functions are related as $\sigma_g \approx 1.29\,\sigma_t$ \cite{sec92}. 
Applying a Kolmogorov-Smirnov (K-S) test (e.g.\ Lindgren 1962), the above 
fit is
only accepted at the 82\% confidence level, indicating that the $t_5$ 
function does not provide an incredibly good fit to the observed GCLF in
the Sombrero over the 
full magnitude range. Without the completeness correction the peak would
be at $\vto=22.22\pm0.12$, while we get $\vto=22.46\pm0.14$ if both the 
completeness and contamination corrections are omitted. Thus, the correction 
for incompleteness only has a small effect on the turn-over magnitude, while 
the contamination correction has a more pronounced effect.

  If the fit is restricted to a narrower magnitude range then the effects 
of contamination, incompleteness and deviations from the assumed $t_5$
shape of the GCLF may be further reduced. If we adopt a magnitude limit
of $V=23.5$ instead of 24.0 then the resulting $t_5$ fit has a turn-over at 
$\vto=22.05\pm0.10$ and $\sigma_t=0.81\pm0.10$, corresponding to a Gaussian
$\sigma_g=1.05$. The K-S test now accepts the fit at the 99.8\% confidence 
level, indicating no statistically significant deviations from a $t_5$
function for this brighter subsample. Leaving out the completeness correction 
has virtually no effect, changing $\vto$ by only 0.01 mag to $22.04\pm0.10$. 
This is not unexpected, considering that the applied magnitude limit is
$\sim 1$ mag brighter than the 50\% completeness limit.  Again, the 
contamination correction has a stronger effect and without it 
the turn-over would be at $\vto=22.13\pm0.12$.  For an even brighter 
magnitude cut-off one expects the fits to become more uncertain as the
sample size decreases and the dispersion becomes less well constrained, 
especially if the dispersion and turn-over are fitted simultaneously. 
Nevertheless, we find that the turn-over magnitude and dispersion of 
the $t_5$ function remain stable for a magnitude cut-off as bright as 22.0, 
even though the formal errors become larger. For cut-offs at $V=23.0$, 
22.5 and 22.0 we find $\vto = 22.10\pm0.12$, $22.10\pm0.18$ and 
$22.08\pm0.24$, respectively.

  Another possibility is to keep the dispersion fixed and only allow
the turn-over to vary. If we adopt a fixed $\sigma_t = 1.0$, corresponding
to a Gaussian $\sigma_G = 1.3$ (e.g. Kundu \& Whitmore \shortcite{kundu01})
and perform a fit down to $V=23.5$ then we get a somewhat fainter turn-over
at $\vto=22.15\pm0.13$. However, the K-S test only accepts this fit at
the 66\% confidence level, indicating a much worse match to the data than for 
the two-parameter fit. 

  We thus adopt $\vto = 22.10\pm0.15$ as our best estimate of the 
reddening-corrected $V$-band turn-over magnitude for the GCLF of the 
Sombrero, where the $\pm0.15$ uncertainty is based on the 1$\sigma$
errors returned by the maximum-likelihood fits as well as the scatter
in the various fits.  This corresponds to an absolute value of 
$\mto = -7.60\pm0.15$ for a distance modulus of 29.7. For comparison, 
Secker \shortcite{sec92} found $\mto = -7.29\pm0.13$ and $\mto = -7.51\pm0.15$
for the Milky Way and M31. Note that the errors on these numbers are
quite similar to our formal errors, consistent with the fact that a
similar number of clusters are fitted.  More recently, Barmby, Huchra and 
Brodie \shortcite{barm01} found a $V$-band turn-over of $\vto = 16.84$ for 
M31, or $\mto=-7.6$. Ferrarese et al.\ \shortcite{fer00} point out that a 
weighted mean to GCLF turn-over magnitudes for galaxies in the 
Virgo and Fornax gives $\mto=-7.60$ with a systematic difference of 0.5
mag between the two galaxy clusters.  Kundu \& Whitmore \shortcite{kundu01} 
favour a value of $\mto=-7.41$ based on a sample of 28 elliptical galaxies, but
do not include any of the Fornax galaxies.  Thus, our turn-over of $-7.60$ 
for the Sombrero may be somewhat on the bright side, but it is not clear how 
much of the scatter in reported turn-over magnitudes from various literature 
sources is intrinsic, and how much of it is due to different measurement 
techniques etc.  The dispersion measured for the Sombrero GCLF 
($\sigma_t = 0.81\pm0.10$ or $\sigma_g=1.05$) is formally somewhat narrower 
than the value in the Milky Way, but agrees well with the one in M31 
($\sigma_g$ = 1.42 and 1.06, respectively, Secker 1992).  

  Overall, we conclude that the GCLF of the Sombrero seems quite similar to 
that in most other well-studied galaxies.  However, it is worth noting that 
the relatively small number of clusters does not provide very strong 
constraints on the exact form of the GCLF, a situation which can only be 
remedied by obtaining a larger sample of GCs.

\begin{figure}
\epsfxsize=85mm
\epsfbox{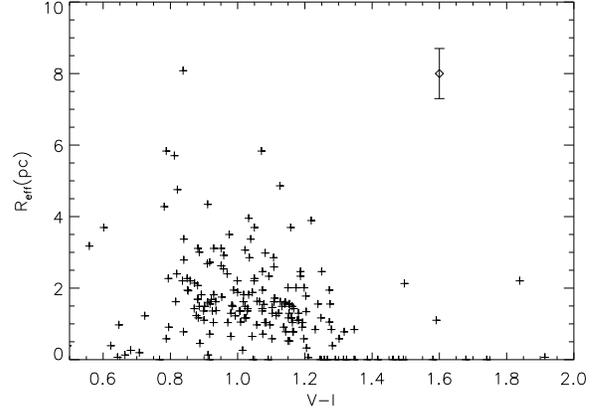}
\caption{\label{fig:vi_fwhm} Cluster half-light radii measured on the 
F547M images (central pointing) as a function of \viz\ colour. The error 
bar indicates the typical random error on individual size measurements,
as estimated in Sect~\ref{sec:gcsz}.
}
\end{figure}

\begin{figure}
\epsfxsize=85mm
\epsfbox{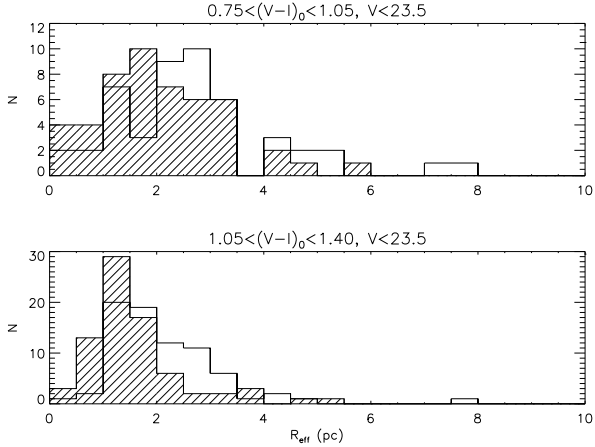}
\caption{\label{fig:szdist}Distributions of half-light radii for blue 
(top) and red (bottom) clusters in M104. On the average, blue clusters are 
larger than red ones.  Hatched and outlined histograms are for sizes measured 
on F547M and F814W band images, respectively.
}
\end{figure}

\begin{figure}
\epsfxsize=85mm
\epsfbox{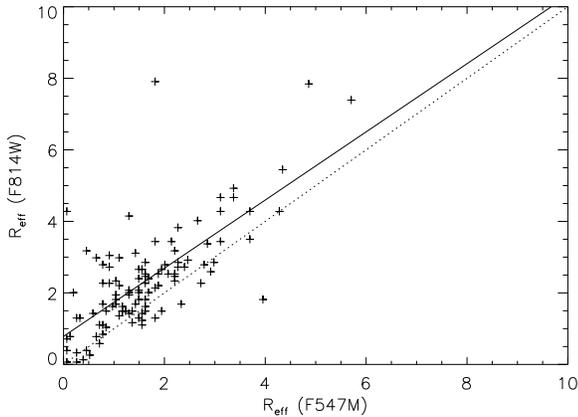}
\caption{\label{fig:sz_v_i} Comparison of object sizes for objects
brighter than $V=23.5$, measured on F547M and M814W images. The solid line
is a fit to the data points while the dashed line represents a
1:1 relation.
}
\end{figure}

\subsubsection{Globular cluster sizes}
\label{sec:gcsz}

  Fig.~\ref{fig:vi_fwhm} shows cluster sizes as a function of \viz\ colour.
The sizes were measured by \ishape\ on $V$ frames, excluding objects
fainter than $V=23.5$ in order to avoid too low S/N objects where the
size information would be uncertain. 

  As can be seen from the figure, the data for GCs in the Sombrero show a 
correlation between cluster size and \viz\ colour similar to that in 
ellipticals and S0 galaxies \cite{kundu98,puz99,lar01}. The size difference 
between blue and red clusters (dividing at $\viz = 1.05$) is seen
somewhat more clearly in Fig.~\ref{fig:szdist} where the size distributions
for blue (top) and red (bottom) clusters are illustrated. In
Fig.~\ref{fig:szdist} the hatched and outlined histograms represent
sizes measured on F547M and F814W band images, respectively. The sizes
measured on the F814W images are generally somewhat larger (see below), but 
the size difference between red and blue GCs is clearly seen in both cases.
The mean effective radii (defined as the radius within which half of
the total cluster luminosity is contained) of red and blue clusters are 
$\reff = 1.61$ pc and $\reff = 2.09$ pc, measured on the F547M images.

  We can also compare with cluster sizes measured on the PC frame. These
should be more accurate, given the better resolution of the PC chip.  For 
the 9 red and 9 blue cluster candidates on the PC chip with size
information, the average sizes are 1.48 and 2.54 pc. Of course, small 
number statistics play a significant role here and the clusters near the 
nucleus might also have different physical sizes from those further out. 
In any case, the size difference between blue and red clusters is 
confirmed by both measurements.

  The accuracy of the measured cluster sizes can be further checked by 
comparing size measurements on the F547M and F814W images. 
Fig.~\ref{fig:sz_v_i} compares sizes measured in the two bands for objects 
brighter than $V=23.5$, the same magnitude limit adopted for the cluster size 
measurements. The dashed line represents a 1:1 relation between the two sets 
of size measurements while the solid line is a least-squares fit to the data. 
As can be seen from the figure, the sizes measured on the F814W exposures are 
on the average $\sim0.75$ pc larger than those measured on the F547M images. 
It is worth noting, though, that one WF camera pixel corresponds to a linear 
scale of 4.2 pc at the adopted distance of the Sombrero.  The 0.75 pc 
offset could be due to small changes in the telescope focus (``breathing''), 
minor offsets between the combined exposures or inaccuracies in the PSF 
modelling by \tinytim .  Until these effects are better understood, the 
\emph{absolute} 
values of the sizes in Fig.~\ref{fig:vi_fwhm} and Fig.~\ref{fig:szdist} 
should not be taken too literally, but size \emph{differences} are much 
more robust. Apart from the 0.75 pc offset, the two sets of size measurements
agree quite well. The standard deviation around the fit indicated by the 
solid line in Fig.~\ref{fig:sz_v_i} is 0.7 pc, indicating that individual 
cluster sizes are accurate at about this level, not counting possible 
systematic effects.

\subsection{The halo pointings}

\begin{figure}
\epsfxsize=85mm
\epsfbox{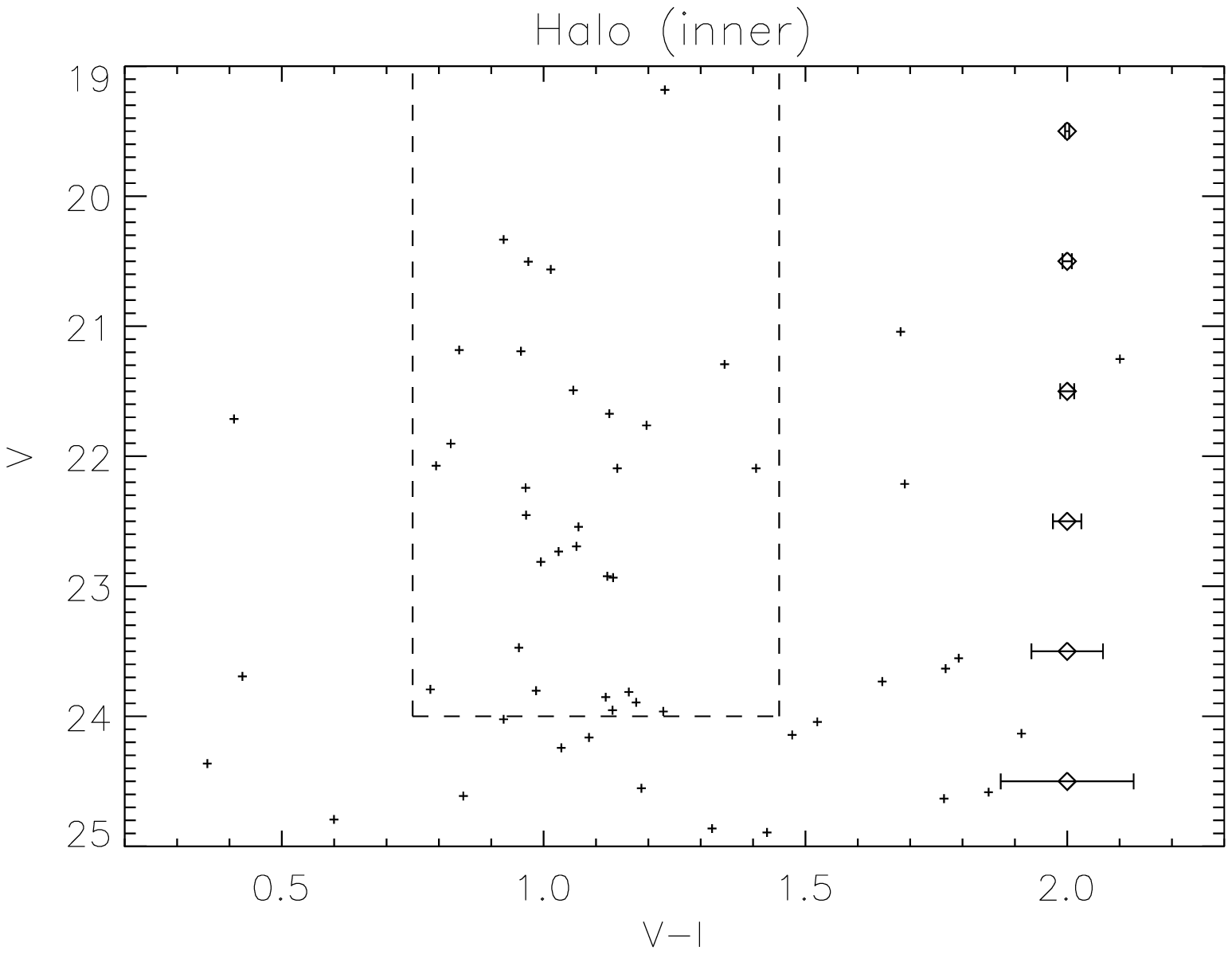}
\epsfxsize=85mm
\epsfbox{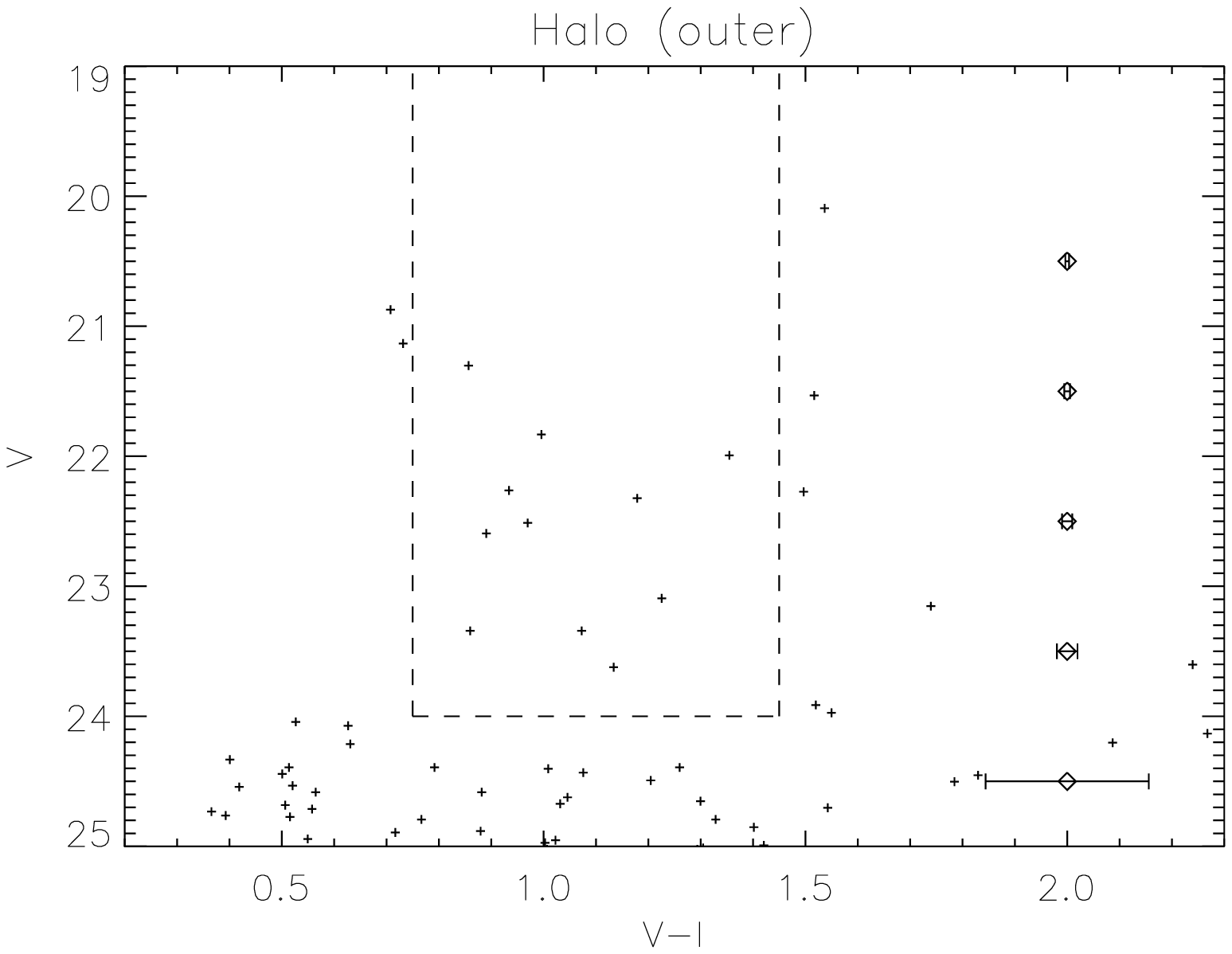}
\caption{\label{fig:cmdh} Colour-magnitude diagram for objects
in the WFPC2 chips in the two halo pointings. 
}
\end{figure}


  The colour-magnitude diagrams for the two halo fields are shown in 
Fig.~\ref{fig:cmdh}. Obviously, the number of GC candidates in the halo 
fields are much smaller than in the central pointing, but these fields 
nevertheless serve as a very useful comparison.  Using the same selection 
criteria as for the central field, the inner and outer halo fields contain 
31 and 11 GC candidates, respectively (Table~\ref{tab:clusters}).  While 
most of the objects in the 
inner halo field are probably true globular clusters, we expect a 
significant fraction of the objects in the outer halo field to be 
contaminants, bearing in mind that about 8 foreground/background objects 
are expected within the GC colour range.

The colour-magnitude diagrams in Fig.~\ref{fig:cmdh} are not quite as 
strikingly bimodal as the corresponding plot for the central field 
(Fig.~\ref{fig:cmd}), and bimodality is only detected at the 40\% confidence 
level by a KMM test when combining data for the two halo pointings. 
Nevertheless, the peaks found by the KMM test are at $\viz = 0.97$ 
and $\viz = 1.19$, nearly identical to those in the central pointing. 
60\% of the objects in the halo pointings are assigned to the blue peak by
the KMM test, compared to 45\% in the central pointing, providing a hint that 
the ratio of blue (metal-poor) to red (metal-rich) clusters increases somewhat
as a function of galactocentric distance.


\subsection{Radial trends}
\label{sec:rad}

\begin{figure}
\epsfxsize=85mm
\epsfbox{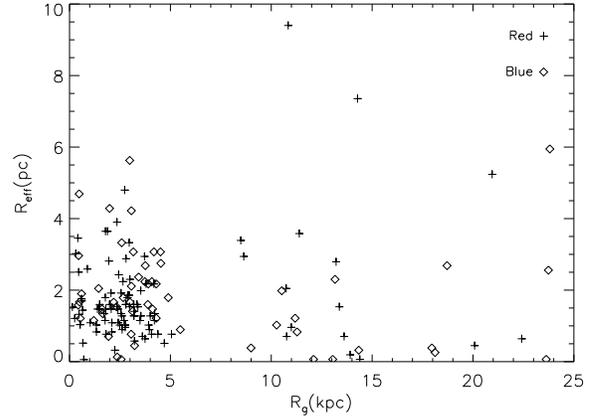}
\caption{\label{fig:reff_d} GC sizes as a function of projected
galactocentric distance in kpc. Plus ($+$) markers indicate red 
($\viz>1.05$) clusters, diamonds ($\diamond$) blue clusters.
}
\end{figure}

\begin{figure}
\epsfxsize=85mm
\epsfbox{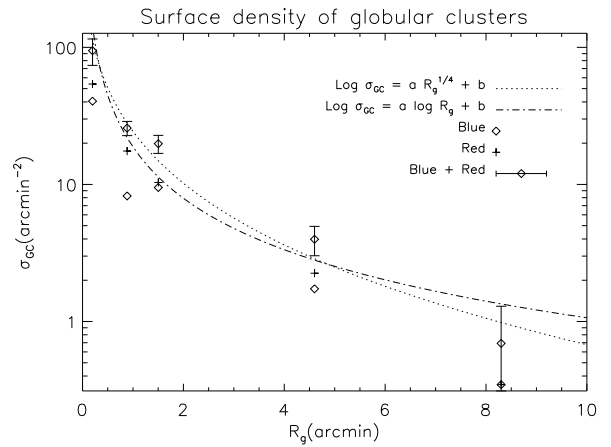}
\caption{\label{fig:rprof} The surface density profile of the Sombrero
GCS. 
The two outer points are for each of
the halo pointings. The two lines represent analytical fits to an
$R^{1/4}$ law and a power law as discussed in the text. One arcmin
equals 2.5 kpc.
}
\end{figure}

  Data for all the three pointings have been combined to produce
Fig.~\ref{fig:reff_d}, which shows cluster sizes as a function of
projected galactocentric distance $R_{\rm g}$. Red ($\viz>1.05$) and 
blue ($\viz<1.05$) clusters are shown with different symbols.  We have 
not attempted to remove contaminants from Fig.~\ref{fig:reff_d}. Although 
some of these will be foreground stars and thus appear point-like, others 
may be background objects with finite sizes that could be in 
the apparent GC size range. In addition to the contamination problem, 
observed radial trends in the cluster sizes may also be affected by the 
fact that the different exposures are taken through different passbands, 
with different exposure times and at different epochs. In particular, 
Holtzman et al.\ \shortcite{hol99} noted that the HST point spread function 
can exhibit significant temporal variation due to the so-called 
``breathing'' of the HST secondary mirror. 

  With these precautions in mind, the plot 
shows little, if any, correlation between cluster size and $R_{\rm g}$,
at least for $R_{\rm g} \la 15$ kpc where most of the objects may
still be expected to be globular clusters.  Indeed, the average effective 
radii of objects in the three pointings are $1.77$, $1.99$ and $2.02$ pc 
for the central, inner and outer halo fields respectively.

  The three HST pointings also allow us to investigate the radial density
profile of the GCS as shown in Fig.~\ref{fig:rprof}. The central pointing
has here been divided into three bins, plotting data for the PC chip 
separately and dividing the WF chip data at 3 kpc or about $1\farcm2$.  
The two halo pointings each contribute with one data point. The total 
surface density of blue$+$red clusters is shown with error bars 
corresponding to Poisson statistics. We also plot data for red and blue 
clusters separately, although the error bars are omitted for 
clarity. The data in Fig.~\ref{fig:rprof} have been corrected for
foreground/background objects by subtracting 4 red and 4 blue objects
from each pointing.

  The azimuthal coverage of our data is clearly limited, but our four
outermost data points for the surface density of globular clusters 
nevertheless agree quite well with results by Bridges \& Hanes 
\shortcite{bri92} and with the GC counts quoted by Harris et 
al.~\shortcite{har84}. The earlier ground-based studies were unable
to probe the central few arcseconds of the Sombrero GCS, covered by
the PC camera. It should, however, be pointed out that the PC data may still 
be uncertain because of possible obscuration from the dust lane
which covers $\sim50$\% of the PC chip, even though we have attempted 
to mask out the most prominent parts of the dust lane.

  We fitted the GC surface density $\sigma_{\rm GC}$ as a function of
galactocentric radius $R_g$ with two different model profiles:
A single power-law fit of the form
\begin{equation}
  \sigma_{\rm GC} \propto R_g^\alpha
  \label{eq:plaw}
\end{equation}
yields an exponent of $\alpha = -1.25\pm0.20$, somewhat steeper than
the value $-0.75\pm0.15$ found by Bridges and Hanes \shortcite{bri92}
for the central $4\arcmin$ but considerably shallower than the
$-1.82$ value found by Harris et al.~\shortcite{har84} for the
outer regions. The actual profile is likely to be better represented
as a composite of two power-laws \cite{bri92}, although our relatively
sparse coverage does not allow a more detailed investigation of this
issue.  However, the data do appear to be better 
fitted by a de Vaucouleurs $R^{1/4}$ profile of the form
\begin{equation}
  \log \sigma_{\rm GC} = \alpha R_g^{1/4} + \beta 
  \label{eq:devau}
\end{equation}
with $\alpha = -2.00\pm0.18$ and $\beta = 3.39\pm0.23$. This is only
slightly shallower than the value $\alpha = -2.12$ found by Harris et al. 
\shortcite{har84}. 

  The total number of GCs in the Sombrero can be estimated by integration
of (\ref{eq:devau}). Adopting an outer radius of $20\arcmin$, 
we thus obtain a total of 1150 clusters which is nearly a factor
of two smaller than the $\sim 2000$ GCs usually thought to exist
in the Sombrero \cite{har84,bri92,for97a}. However, the uncertainties on
the fitted $\alpha$ and $\beta$ values translate to an uncertainty of
about $\pm50$\% on the total number of clusters. The number of clusters
also depends on the completeness correction, as well as on the adopted 
outer radius of the system. If this is taken to be $30\arcmin$ instead of
$20\arcmin$ then the total number of GCs is $\sim 1290$.  We have not
corrected for incomplete sampling of the GC population due to the
magnitude cut-off at V=24. Integrating a $t_5$ function with $\vto=22.1$
and $\sigma_t=0.81$ over the $V=19.0-24.0$ range, one finds that only
about 4\% of the globular clusters would fall outside this range.
Alternatively, one can compare directly with the observed Milky Way
GCLF: For a distance modulus of 29.7 our cut-off at $V=24$ corresponds
to $M_V=-5.7$. Of the 144 GCs in the McMaster catalogue \cite{har96} with
tabulated $M_V$ values, 25 GCs or 17\% fall below this limit. Thus, we
estimate that we may have lost between 4\% and 17\% of the GCs in the 
Sombrero because of the magnitude limit.

  The RC3 catalogue lists a total reddening-corrected $B$ magnitude of 8.38 
and $\bvz = 0.84$ for the Sombrero, leading to an absolute $V$ band
magnitude of $M_V=-22.16$ for distance modulus 29.7.  1150 clusters then 
correspond to a specific frequency of $S_N = 1.6 \pm 0.8$ (where the $\pm0.8$
comes from the $\pm50$\% estimated above), in reasonable agreement with 
earlier results but somewhat on the low side.  Estimates of the bulge 
vs.\ total luminosity for the Sombrero range between 0.73 \cite{bag98} and 
0.85 \cite{ken88}. If we adopt 0.8 as a compromise, the bulge has 
$M_V = -21.92$ and the specific frequency with respect to the bulge alone 
becomes $S_N({\rm bulge}) = 2.0\pm1.0$.  If as many as 17\% of the
GCs are fainter than $V=24$ that would increase $S_N({\rm total})$ to 
around $1.9\pm0.9$ and $S_N({\rm bulge})$ to $2.3\pm1.2$.

  If the red and blue cluster populations are fitted separately we obtain
$\alpha_R = -2.07\pm0.19$ and $\alpha_B = -1.91\pm0.21$, again indicating
that the spatial distribution of blue clusters is probably more extended 
than that of the red ones. The $\alpha$ values can be
converted to effective radii (containing half the total number of
clusters) for the de Vaucouleurs profiles, yielding
$\reff_{,R} = 6\farcm7 \pm 2\farcm6$ and 
$\reff_{,B} = 9\farcm2 \pm 3\farcm9$ for the red and blue GCs
respectively, assuming that the density profile continues
to infinite radius. If we instead adopt an outer radius of $20\arcmin$ as
above, which is probably a more reasonable assumption, then the respective 
effective radii become $\reff_{,R} = 4\farcm5 \pm 1\farcm0$ (11.4 kpc)and 
$\reff_{,B} = 5\farcm4 \pm 1\farcm1$ (13.7 kpc) for the red and blue GCs.

\section{Discussion}

\noindent

  We have clearly detected two sub-populations of GCs in the Sombrero galaxy, 
hinted at in Forbes et al. \shortcite{for97a}. Thus, the Sombrero is like 
most large galaxies in this regard. The blue sub-population, with $(V-I)_0$ 
= 0.96 is quite similar in colour to the halo GCs in our Galaxy and M31, i.e. 
$\viz \sim 0.92$ \cite{barm00}. It is also remarkably similar to the mean 
value for a sample of early-type galaxies studied by Forbes \& Forte
\shortcite{ff01}, i.e. $(V-I)_0$ = 0.954 $\pm$ 0.008. The metallicity
of [Fe/H] $= -1.4$ corresponding to the blue peak is also similar to that
of metal-poor halo clusters in our Galaxy ([Fe/H] $= -1.59$) and M31 
([Fe/H] $= -1.40$) \cite{for00}.  It is therefore likely that the blue GCs 
in all these systems are old and metal--poor. 

  The red GCs have $(V-I)_0$ = 1.21 or [Fe/H] $= -0.54$.  This is very 
similar to the metallicities of the ``metal-rich'' sub-population in both 
our Galaxy and M31, i.e. [Fe/H] $ = -0.55$ and $-0.58$ respectively 
\cite{for00}. So again, by analogy with known spirals, we associate the 
red GCs in the Sombrero with a bulge/disk population. 

\begin{figure}
\epsfxsize=85mm
\epsfbox{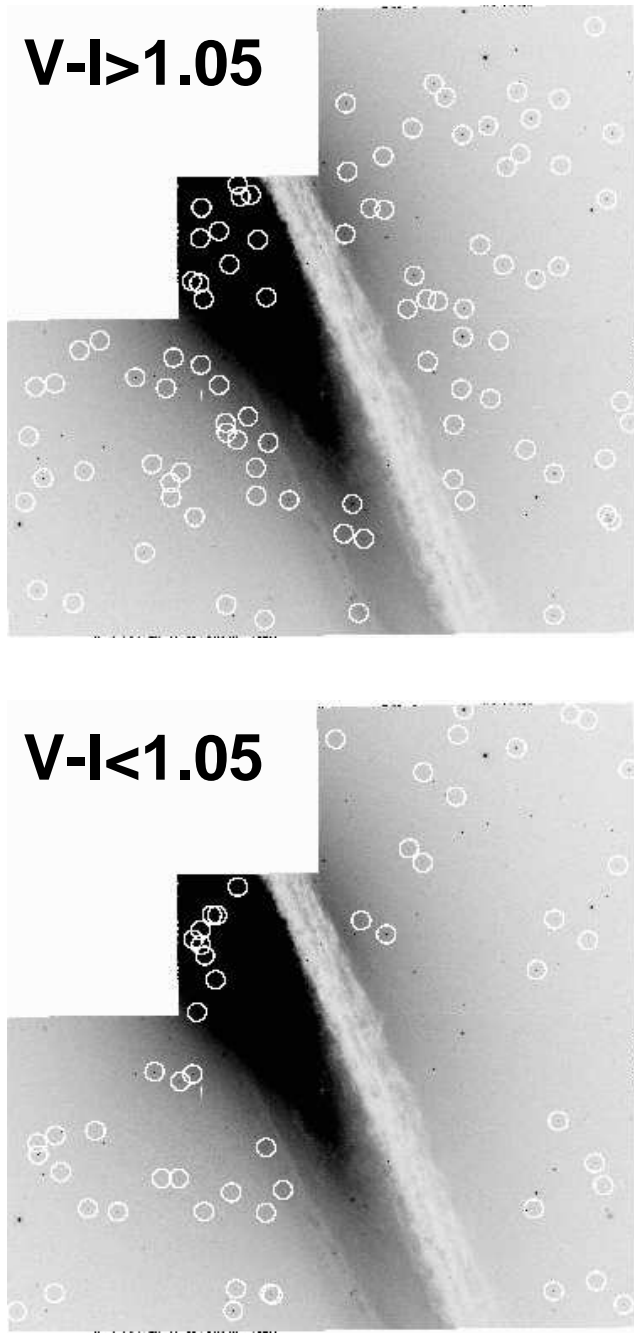}
\caption{\label{fig:spat}Spatial distributions of red (top) and blue
(bottom) clusters. The dust lane and the brightest parts of the
disk were masked out before photometry, so the absence of objects
near the disk is not real.  The apparent peculiar alignment of blue objects 
in the PC frame is probably just a consequence of the small number
of detections in this chip.
}
\end{figure}

  The relation between host galaxy luminosity and the colours of globular 
cluster subpopulations has been studied in a number of recent papers.
Forbes et al.\ \shortcite{for97b} found a correlation between host galaxy
luminosity and the colour/metallicity of the red GCs, but no significant
trend with luminosity for the blue GCs. More recently, Larsen et al. 
\shortcite{lar01} found correlations at the $2-3\sigma$ level with host 
galaxy luminosity for the colours of \emph{both} GC populations.
Kundu \& Whitmore \shortcite{kundu01} and Burgarella, Kissler-Patig \& Buat
\shortcite{bur01} found at best weak correlations between host galaxy 
luminosity and GC colours, while Forbes \& Forte \shortcite{ff01} detected
a correlation between the colour of the red GCs and host galaxy central
velocity dispersion, but no correlation for the blue GCs.  The Sombrero does 
not represent an outlier with respect to any of the above datasets and was, 
in fact, included in the samples studied by both Larsen et al. 
\shortcite{lar01} and Burgarella, Kissler-Patig \& Buat \shortcite{bur01}. 
Note, however, that the mean colours of the two GC populations quoted here 
are slightly different from those in Larsen et al. \shortcite{lar01} (0.939 
and 1.184 instead of the 0.96 and 1.21 values in this paper), even though
the two studies use the same dataset. This difference 
reflects the slightly different colour / magnitude cuts adopted in the two 
analyses, but is within our estimated $\pm0.03$ mag error margins. In fact, 
with the GC colours in the present paper the Sombrero fits even better onto 
the Larsen et al. \shortcite{lar01} relation for galaxy luminosity 
vs.\ GC colour.

  In spiral galaxies the blue GCs are naturally associated with a halo 
population. In our Galaxy, the metal-poor halo GCs are distributed 
nearly spherically around the galaxy centre while the metal-rich population 
has a somewhat flattened distribution \cite{kin59,zin80}.  The spatial 
distributions of red and blue clusters in the central Sombrero pointing are 
shown in Fig.~\ref{fig:spat}. From this figure it seems that the spatial 
distributions of red and blue GCs are not strikingly different within 
the central $\sim 4$ kpc covered by the archive images, but the spatial
distribution of the red GCs is clearly different from the stellar disk
of the Sombrero.  The limited spatial coverage of WFPC2 makes it difficult 
to draw firm conclusions regarding any differences in the distributions of 
red and blue clusters further out. However, there are some indications that 
the number of blue clusters declines more slowly as a function of distance 
from the centre, implying that the ratio of blue to red clusters increases 
outward. 

  An interesting difference between the Sombrero GC system and that of 
other spiral galaxies is the relatively low ratio of metal-poor (blue) to
metal-rich (red) GCs, i.e. about 0.8:1 in the central pointing and
perhaps slightly higher for the GCS as a whole.  Using metallicities
and galactocentric radii for GCs in the Milky Way from the McMaster 
catalogue \cite{har96} and dividing between metal-rich and metal-poor 
clusters at [Fe/H] $= -1.0$, the corresponding ratio within the central 
4 kpc of our own Galaxy is $\sim 1.2:1$. In M31 it is $\sim 1.0:1$, using 
data from Barmby et al. \shortcite{barm00}. Further out in the halo of
spiral galaxies metal-poor clusters dominate, leading to global 
blue-to-red ratios of about 2:1 and 3:1 in the Milky Way and M31 respectively.
For early type galaxies the ratio is almost always $>$1:1 although
red GCs also here tend to be more centrally concentrated than blue ones
\cite{for97b}.

  Given that the stellar bulge-to-disk ratio for Sombrero is 6:1 \cite{ken88}
and that the overall ratio of metal-poor to metal-rich globular clusters
is much lower than in spiral galaxies with less conspicuous bulges, it seems 
very unlikely that the red GCs are associated with the disk but must rather 
be associated with the bulge.  If the red GCs were associated with the disk 
we might expect a number roughly similar to that in the Milky Way, i.e.\ 
about 50 or so.  In Section~\ref{sec:rad} we estimated a total number of 
around 1150 GCs in the Sombrero so if there are roughly equal numbers of 
red and blue GCs then that would imply a total of $\sim600$ red GCs.  Thus 
the situation in the Sombrero lends support to the idea that the metal-rich 
GCs also in our own and other spiral galaxies are more closely associated 
with the bulge than the disk \cite{har76,fre82,min95}. 

  Another similarity between the Sombrero and other spirals, as well as 
early-type galaxies, is the relationship between GC colour and physical size. 
Here we find that the blue GCs are $\sim$30\% larger than the red ones. This 
is seen in several early type galaxies and also in the Milky Way \cite{lar01}. 
It is not currently clear what causes this difference, which could be either 
set up at formation e.g. because of different conditions in the globular 
cluster progenitor clouds, or due to ongoing dynamical processes.  
As noted above, the blue GC population does appear to be more spatially 
extended than the red, but the current data are too sparse to study
radial trends in the cluster sizes in detail and e.g.\ check how the GC sizes
correlate with galactocentric distance. It is also possible that the
different GC populations are on different orbits, in which case dynamical
destruction processes might also affect them in different ways.

\section{Concluding Remarks}

Using HST imaging of the central few kpc of the nearby Sombrero galaxy we 
have detected $\sim$150 globular cluster candidates.  We find a  bimodal GC 
colour distribution with peaks at $\viz = 0.96$ and 1.21, corresponding 
to [Fe/H] = $-1.4$ and $-0.54$ \cite{kis98}. The blue GCs have a mean size 
of 2.09 and 
the reds 1.61 pc. Including additional data for two halo pointings at
projected distances of 11 kpc and 21 kpc from the nucleus of the Sombrero,
we have studied radial trends in cluster sizes and the relative numbers
of red and blue clusters. We find that the red clusters appear to be
more centrally concentrated in the galaxy than the blue ones, although 
it would be highly desirable to confirm this result with better sample 
statistics.  As an Sa galaxy, the Sombrero offers a potential link between 
the GC systems of elliptical galaxies and later type spirals.  Indeed we 
find several similarities between the GC system of the Sombrero and galaxies 
of all types. \\

\noindent
$\bullet$ The blue GCs have a mean colour (and hence metallicity) that 
is similar to those seen both in large spirals and ellipticals. \\

\noindent
$\bullet$ The size-colour relationship is qualitatively the same as that
seen in the Milky Way and several well-observed ellipticals. \\

\noindent
$\bullet$ The red GCs have an inferred metallicity that is very close to the 
metal-rich populations of the Milky Way and M31 and other large galaxies.\\

\noindent
$\bullet$ Without corrections for incomplete sampling of the GCLF,
the GC specific frequency normalised to the total galaxy luminosity
is $S_N({\rm total}) = 1.6\pm0.8$, or $S_N({\rm bulge}) = 2.0\pm1.0$
when normalised to the bulge. Corrections for incomplete sampling would
increase these numbers by between 4\% and 17\%. The specific frequency
of the Sombrero is thus similar to that of field and group ellipticals. \\

\noindent
$\bullet$ The large number of red GCs compared to other spirals, combined
with the high bulge/disk ratio in the Sombrero and the spatial 
distribution of the red GCs, suggests that this GC sub-population is
associated with the bulge (rather than the disk) component. \\

\noindent
The proximity and rich GC system of the Sombrero galaxy clearly warrant 
further study. Better HST coverage beyond the central few kpc would improve 
sample statistics and size/colour trends with galactocentric radius could 
be better probed. Spectra with $8 - 10$ m class telescopes would provide 
direct metallicity and age estimates for individual GCs. Kinematics of 
the two subpopulations could also be explored. 

\section{Acknowledgements}

  This work was supported by National Science Foundation grant number
AST9900732, Faculty Research funds from the University of California,
Santa Cruz and NATO Collaborative Research grant CRG 971552. We are
grateful to Carl Grillmair and an anonymous referee for their useful 
comments. DF thanks the hospitality of Lick Observatory where much of 
this work was carried out.

\begin{table}
\caption{\label{tab:clusters}
 Data for clusters in the Sombrero. $V$ and \viz\ have been corrected for 
 a foreground extinction of $A_V = 0.17$ and $E(V-I)=0.086$ mag. The
 last column is the estimated effective (half-light) cluster radius in pc.
 Note that sizes for objects fainter than $V=23.5$ (in parentheses) are 
 uncertain.  Prefixes `C', `H1' and `H2' denote objects in the central, outer 
 and inner halo pointings, respectively.
}
\begin{tabular}{rrrrrr}
  ID & RA(2000.0) & DEC(2000.0) & $V$ & \viz & \reff \\ \hline
C-001 & 12:40:00.31 & -11:37:21.4 & 19.22 &  1.28 & 1.78 \\
C-002 & 12:39:59.40 & -11:37:25.2 & 20.88 &  1.15 & 1.55 \\
C-003 & 12:39:58.69 & -11:37:28.2 & 21.53 &  1.08 & 1.05 \\
C-004 & 12:39:58.65 & -11:37:25.8 & 20.18 &  1.34 & 1.05 \\
C-005 & 12:39:58.52 & -11:37:25.9 & 22.85 &  0.90 & 1.92 \\
C-006 & 12:39:58.45 & -11:37:26.5 & 20.70 &  1.17 & 1.46 \\
C-007 & 12:39:59.58 & -11:37:16.7 & 21.21 &  1.07 & 1.22 \\
C-008 & 12:39:59.02 & -11:37:17.8 & 20.14 &  1.09 & 3.06 \\
C-009 & 12:39:58.81 & -11:37:18.7 & 22.54 &  1.04 & 3.50 \\
C-010 & 12:39:59.76 & -11:37:11.8 & 21.47 &  0.95 & 1.69 \\
C-011 & 12:39:58.77 & -11:37:17.6 & 20.11 &  1.03 & 2.54 \\
C-012 & 12:39:59.33 & -11:37:12.0 & 20.96 &  0.98 & 1.60 \\
C-013 & 12:39:59.92 & -11:37:07.7 & 23.31 &  1.27 & 0.06 \\
C-014 & 12:39:58.55 & -11:37:16.4 & 21.52 &  1.17 & 1.72 \\
C-015 & 12:39:59.15 & -11:37:12.3 & 22.45 &  0.95 & 3.00 \\
C-016 & 12:39:58.92 & -11:37:13.6 & 20.79 &  0.89 & 4.75 \\
C-017 & 12:39:59.00 & -11:37:13.0 & 23.08 &  1.26 & 1.34 \\
C-018 & 12:39:59.66 & -11:37:08.4 & 23.80 &  1.11 & (err) \\
C-019 & 12:40:00.12 & -11:37:04.6 & 22.25 &  1.02 & 2.62 \\
C-020 & 12:39:59.01 & -11:37:11.1 & 21.96 &  0.78 & 1.22 \\
C-021 & 12:39:59.58 & -11:37:06.9 & 23.22 &  1.19 & 0.52 \\
C-022 & 12:39:59.85 & -11:36:41.3 & 22.58 &  1.24 & 3.69 \\
C-023 & 12:39:59.86 & -11:36:36.0 & 23.92 &  1.39 & (2.01) \\
C-024 & 12:40:00.59 & -11:36:55.4 & 21.70 &  1.24 & err \\
C-025 & 12:40:00.59 & -11:36:55.4 & 21.70 &  1.24 & 1.49 \\
C-026 & 12:40:00.88 & -11:37:00.6 & 22.60 &  1.31 & 0.84 \\
C-027 & 12:40:00.73 & -11:36:49.6 & 21.02 &  0.98 & 1.36 \\
C-028 & 12:40:00.63 & -11:36:45.3 & 20.83 &  1.10 & 1.81 \\
C-029 & 12:40:01.02 & -11:36:57.5 & 21.94 &  1.01 & 1.62 \\
C-030 & 12:40:00.18 & -11:36:27.4 & 22.73 &  1.25 & 1.10 \\
C-031 & 12:40:01.30 & -11:37:02.4 & 23.58 &  1.19 & (2.85) \\
C-032 & 12:40:01.05 & -11:36:54.1 & 20.38 &  0.97 & 1.49 \\
C-033 & 12:40:00.10 & -11:36:23.0 & 22.60 &  1.25 & 1.30 \\
C-034 & 12:40:01.00 & -11:36:50.8 & 22.32 &  1.15 & 1.36 \\
C-035 & 12:40:01.96 & -11:37:05.5 & 21.70 &  1.24 & 0.78 \\
C-036 & 12:40:01.90 & -11:37:00.0 & 22.99 &  1.13 & 3.69 \\
C-037 & 12:40:02.06 & -11:36:59.3 & 21.92 &  1.20 & 1.62 \\
C-038 & 12:40:01.22 & -11:36:31.0 & 23.87 &  0.83 & (1.17) \\
C-039 & 12:40:00.79 & -11:36:16.4 & 22.22 &  1.22 & 1.56 \\
C-040 & 12:40:01.02 & -11:36:21.9 & 22.74 &  1.03 & 2.92 \\
C-041 & 12:40:02.24 & -11:37:00.8 & 22.55 &  1.35 & 1.94 \\
C-042 & 12:40:02.49 & -11:37:07.3 & 21.19 &  1.26 & 1.10 \\
C-043 & 12:40:01.01 & -11:36:17.3 & 22.34 &  0.88 & 5.70 \\
C-044 & 12:40:02.60 & -11:37:05.9 & 21.37 &  1.00 & 1.68 \\
C-045 & 12:40:01.19 & -11:36:16.3 & 21.32 &  0.94 & 2.14 \\
C-046 & 12:40:02.03 & -11:36:40.2 & 23.44 &  1.25 & 5.31 \\
C-047 & 12:40:02.77 & -11:37:02.0 & 22.65 &  1.11 & 3.95 \\
C-048 & 12:40:01.68 & -11:36:24.9 & 22.64 &  1.42 & 11.66 \\
C-049 & 12:40:02.34 & -11:36:45.6 & 23.69 &  1.28 & (0.78) \\
C-050 & 12:40:01.60 & -11:36:19.5 & 22.51 &  0.90 & 0.78 \\
C-051 & 12:40:01.51 & -11:36:15.5 & 21.17 &  1.18 & 1.62 \\
C-052 & 12:40:02.37 & -11:36:40.5 & 23.33 &  0.97 & 0.13 \\
C-053 & 12:40:02.48 & -11:36:44.0 & 21.46 &  0.91 & 3.37 \\
C-054 & 12:40:02.43 & -11:36:42.3 & 22.83 &  1.26 & 0.91 \\
C-055 & 12:40:03.35 & -11:37:05.2 & 23.06 &  1.00 & 1.81 \\
C-056 & 12:40:03.19 & -11:36:59.4 & 23.83 &  1.34 & (6.16) \\
C-057 & 12:40:03.04 & -11:36:53.9 & 23.45 &  1.00 & err \\
C-058 & 12:40:03.04 & -11:36:53.9 & 23.45 &  0.99 & 1.04 \\
C-059 & 12:40:02.65 & -11:36:40.9 & 21.14 &  1.17 & 0.97 \\
C-060 & 12:40:03.48 & -11:37:06.0 & 22.06 &  1.20 & 4.86 \\
\end{tabular}
\end{table}

\begin{table}
Table~\ref{tab:clusters} (continued) \\
\begin{tabular}{rrrrrr} \\ \hline
C-061 & 12:40:01.74 & -11:36:09.2 & 22.18 &  1.08 & 1.17 \\
C-062 & 12:40:02.32 & -11:36:21.6 & 21.11 &  0.95 & 0.45 \\
C-063 & 12:40:03.15 & -11:36:46.1 & 23.90 &  1.02 & (0.32) \\
C-064 & 12:40:02.58 & -11:36:27.7 & 21.10 &  0.94 & 1.43 \\
C-065 & 12:40:03.58 & -11:36:59.3 & 23.65 &  1.04 & (1.04) \\
C-066 & 12:40:03.10 & -11:36:44.0 & 22.67 &  1.07 & 1.88 \\
C-067 & 12:40:03.30 & -11:36:29.7 & 22.19 &  1.22 & 1.56 \\
C-068 & 12:40:03.13 & -11:36:03.1 & 20.37 &  1.11 & 2.20 \\
C-069 & 12:40:04.51 & -11:36:45.3 & 23.57 &  0.87 & (1.68) \\
C-070 & 12:40:04.81 & -11:36:52.1 & 21.35 &  0.92 & 2.27 \\
C-071 & 12:40:03.36 & -11:36:05.9 & 23.95 &  0.81 & (10.89) \\
C-072 & 12:40:04.85 & -11:36:52.4 & 21.85 &  0.99 & 2.72 \\
C-073 & 12:40:03.56 & -11:36:09.5 & 23.56 &  1.22 & (0.84) \\
C-074 & 12:40:04.61 & -11:36:42.1 & 21.56 &  1.06 & 2.20 \\
C-075 & 12:40:04.82 & -11:36:42.4 & 20.91 &  0.90 & 2.20 \\
C-076 & 12:40:03.39 & -11:35:56.0 & 21.36 &  0.90 & 2.79 \\
C-077 & 12:40:05.09 & -11:36:48.7 & 21.42 &  1.20 & 0.91 \\
C-078 & 12:40:04.34 & -11:37:14.4 & 22.21 &  1.18 & 0.58 \\
C-079 & 12:40:03.96 & -11:37:19.1 & 19.88 &  1.12 & 3.37 \\
C-080 & 12:40:05.61 & -11:37:09.6 & 22.66 &  1.39 & 0.78 \\
C-081 & 12:40:04.55 & -11:37:18.1 & 21.73 &  1.14 & 1.62 \\
C-082 & 12:40:02.34 & -11:37:49.4 & 23.10 &  1.28 & 0.32 \\
C-083 & 12:40:04.27 & -11:37:43.2 & 23.52 &  1.38 & (0.97) \\
C-084 & 12:40:03.45 & -11:37:48.8 & 23.07 &  1.10 & 1.81 \\
C-085 & 12:40:02.98 & -11:37:53.8 & 22.75 &  1.10 & 1.17 \\
C-086 & 12:40:02.21 & -11:37:59.4 & 19.57 &  1.27 & 2.46 \\
C-087 & 12:40:04.66 & -11:37:43.5 & 22.43 &  1.13 & 2.01 \\
C-088 & 12:40:03.32 & -11:37:59.1 & 23.31 &  1.33 & 0.39 \\
C-089 & 12:40:02.51 & -11:38:06.8 & 23.23 &  1.09 & 1.62 \\
C-090 & 12:40:04.30 & -11:38:01.7 & 22.52 &  1.14 & 0.71 \\
C-091 & 12:40:05.30 & -11:37:57.1 & 22.16 &  0.96 & 1.49 \\
C-092 & 12:40:06.66 & -11:37:53.0 & 21.41 &  0.95 & 1.81 \\
C-093 & 12:40:06.95 & -11:37:51.1 & 20.93 &  1.21 & 0.78 \\
C-094 & 12:40:04.85 & -11:38:05.4 & 20.21 &  1.05 & 1.30 \\
C-095 & 12:40:04.15 & -11:38:11.1 & 21.56 &  1.05 & 0.65 \\
C-096 & 12:40:05.03 & -11:38:14.7 & 21.92 &  0.93 & 1.23 \\
C-097 & 12:40:06.80 & -11:38:03.2 & 23.39 &  0.81 & 2.79 \\
C-098 & 12:40:05.42 & -11:38:14.3 & 23.04 &  0.99 & 3.11 \\
C-099 & 12:40:04.97 & -11:38:17.8 & 23.95 &  1.12 & (0.58) \\
C-100 & 12:40:05.82 & -11:38:12.5 & 22.43 &  1.22 & 0.52 \\
C-101 & 12:40:05.93 & -11:38:12.9 & 22.32 &  1.31 & err \\
C-102 & 12:40:04.23 & -11:38:26.9 & 22.16 &  1.30 & 1.17 \\
C-103 & 12:40:07.33 & -11:38:06.7 & 22.69 &  0.84 & 0.91 \\
C-104 & 12:40:04.60 & -11:38:26.7 & 21.50 &  1.23 & 0.78 \\
C-105 & 12:40:01.80 & -11:38:02.5 & 21.88 &  1.32 & 1.56 \\
C-106 & 12:40:01.42 & -11:37:50.3 & 23.85 &  1.15 & (0.19) \\
C-107 & 12:40:01.52 & -11:37:57.6 & 22.12 &  1.21 & 1.49 \\
C-108 & 12:40:01.41 & -11:37:55.5 & 22.47 &  1.12 & 2.85 \\
C-109 & 12:40:01.83 & -11:38:20.8 & 22.30 &  1.26 & 2.33 \\
C-110 & 12:40:00.98 & -11:37:55.2 & 20.16 &  1.23 & 1.17 \\
C-111 & 12:40:01.76 & -11:38:21.6 & 21.36 &  0.96 & 1.68 \\
C-112 & 12:40:01.81 & -11:38:26.9 & 22.11 &  1.10 & 1.43 \\
C-113 & 12:40:01.42 & -11:38:15.3 & 22.04 &  1.13 & 2.27 \\
C-114 & 12:40:00.96 & -11:38:12.3 & 21.87 &  1.10 & 1.49 \\
C-115 & 12:40:01.66 & -11:38:35.6 & 23.48 &  1.03 & 12.96 \\
C-116 & 12:40:00.22 & -11:37:53.0 & 19.98 &  0.96 & 2.07 \\
C-117 & 12:39:59.90 & -11:37:44.5 & 22.24 &  1.22 & 1.10 \\
C-118 & 12:40:01.12 & -11:38:30.5 & 23.14 &  0.95 & 3.11 \\
C-119 & 12:39:59.85 & -11:37:49.0 & 22.02 &  0.96 & 1.17 \\
C-120 & 12:39:59.80 & -11:37:55.4 & 21.98 &  1.23 & 1.49 \\
\end{tabular}
\end{table}

\begin{table}
Table~\ref{tab:clusters} (continued) \\ 
\begin{tabular}{rrrrrr} \hline
C-121 & 12:39:59.68 & -11:37:52.4 & 22.20 &  1.17 & 1.04 \\
C-122 & 12:40:01.13 & -11:38:44.0 & 21.03 &  1.15 & 2.98 \\
C-123 & 12:40:00.74 & -11:38:49.5 & 21.49 &  0.88 & 1.62 \\
C-124 & 12:39:58.99 & -11:37:51.2 & 23.34 &  1.18 & err \\
C-125 & 12:39:59.99 & -11:38:25.7 & 21.78 &  1.18 & 1.62 \\
C-126 & 12:40:00.33 & -11:38:37.5 & 23.93 &  1.18 & (12.96) \\
C-127 & 12:39:59.40 & -11:38:08.0 & 21.69 &  0.98 & 4.34 \\
C-128 & 12:39:59.00 & -11:38:00.4 & 23.79 &  1.13 & (err) \\
C-129 & 12:39:59.88 & -11:38:30.0 & 22.53 &  1.11 & 1.88 \\
C-130 & 12:39:59.09 & -11:38:06.5 & 22.81 &  0.98 & 0.71 \\
C-131 & 12:40:00.19 & -11:38:51.9 & 21.26 &  1.26 & 1.04 \\
C-132 & 12:39:59.21 & -11:38:19.5 & 20.91 &  1.15 & 1.04 \\
C-133 & 12:39:58.78 & -11:38:09.5 & 22.43 &  1.35 & 0.84 \\
C-134 & 12:39:59.25 & -11:38:25.8 & 20.29 &  1.24 & 1.04 \\
C-135 & 12:39:59.43 & -11:38:35.8 & 21.18 &  1.22 & 1.30 \\
C-136 & 12:39:57.97 & -11:37:57.7 & 21.53 &  1.06 & 1.49 \\
C-137 & 12:39:59.33 & -11:38:43.8 & 21.10 &  1.19 & 1.30 \\
C-138 & 12:39:58.63 & -11:38:21.7 & 23.85 &  0.84 & (15.55) \\
C-139 & 12:39:58.63 & -11:38:21.7 & 23.85 &  0.84 & (err) \\
C-140 & 12:39:58.54 & -11:38:19.6 & 21.68 &  1.14 & 1.94 \\
C-141 & 12:39:58.94 & -11:38:35.6 & 21.99 &  1.16 & 1.30 \\
C-142 & 12:39:58.26 & -11:38:18.5 & 21.14 &  1.08 & 1.36 \\
C-143 & 12:39:58.05 & -11:38:17.2 & 23.51 &  0.96 & (2.92) \\
C-144 & 12:39:59.38 & -11:39:01.5 & 21.03 &  0.92 & 2.20 \\
C-145 & 12:39:58.31 & -11:38:39.5 & 20.09 &  0.89 & 2.40 \\
C-146 & 12:39:56.98 & -11:38:01.7 & 21.68 &  0.98 & 1.56 \\
C-147 & 12:39:57.72 & -11:38:28.3 & 22.34 &  0.84 & 4.28 \\
C-148 & 12:39:58.48 & -11:38:58.7 & 22.60 &  1.08 & 1.36 \\
C-149 & 12:39:58.37 & -11:38:57.6 & 21.89 &  1.00 & 3.11 \\
C-150 & 12:39:58.16 & -11:38:54.6 & 22.11 &  0.84 & 2.27 \\
C-151 & 12:39:57.39 & -11:38:32.1 & 19.41 &  0.90 & 16.33 \\
\end{tabular}
\end{table}

\begin{table}
Table~\ref{tab:clusters} (continued) \\ 
\begin{tabular}{rrrrrr} \hline
H1-01 & 12:40:20.67 & -11:31:09.2 & 23.62 &  1.13 & (3.65) \\
H1-02 & 12:40:20.31 & -11:31:12.3 & 22.32 &  1.18 & 5.31 \\
H1-03 & 12:40:20.11 & -11:30:26.5 & 21.94 &  1.35 & 0.65 \\
H1-04 & 12:40:21.98 & -11:30:07.4 & 22.54 &  0.89 & 6.03 \\
H1-05 & 12:40:25.72 & -11:30:60.0 & 21.78 &  1.00 & 2.59 \\
H1-06 & 12:40:23.62 & -11:31:40.4 & 23.29 &  1.07 & 6.35 \\
H1-07 & 12:40:25.23 & -11:32:03.3 & 23.29 &  0.86 & 14.71 \\
H1-08 & 12:40:28.03 & -11:31:43.5 & 21.25 &  0.86 & 0.06 \\
H1-09 & 12:40:23.88 & -11:32:37.2 & 23.04 &  1.23 & 0.45 \\
H1-10 & 12:40:19.21 & -11:32:07.0 & 22.46 &  0.97 & 2.72 \\
H1-11 & 12:40:20.67 & -11:32:49.9 & 22.21 &  0.93 & 0.26 \\
H2-01 & 12:40:01.31 & -11:32:17.5 & 21.67 &  1.13 & 2.83 \\
H2-02 & 12:40:02.74 & -11:31:57.5 & 23.91 &  1.23 & (0.45) \\
H2-03 & 12:40:02.91 & -11:32:12.8 & 23.84 &  1.18 & (23.33) \\
H2-04 & 12:40:03.43 & -11:32:23.1 & 23.42 &  0.95 & 2.98 \\
H2-05 & 12:40:04.24 & -11:32:21.1 & 21.71 &  1.20 & 1.56 \\
H2-06 & 12:40:04.49 & -11:31:59.1 & 20.28 &  0.92 & 0.32 \\
H2-07 & 12:40:04.49 & -11:31:57.8 & 22.04 &  1.41 & 0.06 \\
H2-08 & 12:40:04.93 & -11:32:02.2 & 22.88 &  1.13 & 7.45 \\
H2-09 & 12:40:05.55 & -11:32:31.7 & 20.45 &  0.97 & 2.33 \\
H2-10 & 12:40:06.01 & -11:32:15.7 & 20.51 &  1.01 & 0.19 \\
H2-11 & 12:40:06.15 & -11:32:30.2 & 23.74 &  0.78 & (0.52) \\
H2-12 & 12:40:06.40 & -11:32:38.9 & 21.85 &  0.82 & 0.06 \\
H2-13 & 12:40:06.75 & -11:32:26.9 & 21.24 &  1.35 & 0.71 \\
H2-14 & 12:40:03.55 & -11:33:10.8 & 22.76 &  0.99 & 1.23 \\
H2-15 & 12:40:07.53 & -11:33:10.0 & 22.02 &  0.80 & 0.06 \\
H2-16 & 12:40:05.53 & -11:33:18.0 & 23.97 &  0.92 & (8.49) \\
H2-17 & 12:40:06.14 & -11:33:21.3 & 22.40 &  0.97 & 0.84 \\
H2-18 & 12:40:07.24 & -11:33:26.2 & 22.49 &  1.07 & 3.63 \\
H2-19 & 12:40:03.00 & -11:33:30.6 & 22.19 &  0.97 & 1.04 \\
H2-20 & 12:40:06.34 & -11:33:29.9 & 21.44 &  1.06 & 0.97 \\
H2-21 & 12:40:04.39 & -11:33:36.1 & 23.75 &  0.99 & (17.50) \\
H2-22 & 12:40:02.07 & -11:33:16.2 & 19.13 &  1.23 & 0.71 \\
H2-23 & 12:40:01.21 & -11:33:12.6 & 22.68 &  1.03 & 9.53 \\
H2-24 & 12:40:01.02 & -11:33:35.9 & 23.76 &  1.16 & (3.37) \\
H2-25 & 12:40:00.51 & -11:33:04.9 & 23.80 &  1.12 & (err) \\
H2-26 & 12:40:00.45 & -11:34:02.7 & 22.87 &  1.12 & 2.98 \\
H2-27 & 12:40:00.09 & -11:34:05.8 & 22.04 &  1.14 & 3.43 \\
H2-28 & 12:39:59.79 & -11:33:00.5 & 23.90 &  1.13 & (8.81) \\
H2-29 & 12:39:58.63 & -11:33:13.7 & 22.64 &  1.06 & 2.07 \\
H2-30 & 12:39:58.04 & -11:33:19.4 & 21.14 &  0.96 & 2.01 \\
H2-31 & 12:39:57.88 & -11:33:55.0 & 21.13 &  0.84 & 0.39 \\
\end{tabular}
\end{table}

\end{document}